\documentclass[aps,pre,twocolumn,superscriptaddress,showpacs]{revtex4}
\usepackage{amsfonts,amssymb,amsmath,latexsym,epsfig} 
\begin{document}

\title{Irreversible Opinion Spreading on Scale-Free Networks}

\author{Juli\'an Candia}
\affiliation{Consortium of the Americas for Interdisciplinary Science and
Department of\\ Physics and Astronomy,
University of New Mexico, Albuquerque, NM 87131, USA}
\affiliation{Center for Complex Network Research and Department of Physics,\\
University of Notre Dame, Notre Dame, IN 46556, USA}

\begin{abstract}
We study the dynamical and critical behavior of a model for irreversible 
opinion spreading on Barab\'asi-Albert (BA) scale-free networks by performing 
extensive Monte Carlo simulations. The opinion spreading within an inhomogeneous society 
is investigated by means of the magnetic Eden model, a nonequilibrium kinetic model 
for the growth of binary mixtures in contact with a thermal bath. 
The deposition dynamics, which is studied as a function of the degree of the occupied 
sites, shows evidence for the leading role played by hubs in the growth process. 
Systems of finite size grow either ordered or disordered, depending on the 
temperature. By means of standard finite-size scaling procedures,
the effective order-disorder phase transitions are found to persist in the  
thermodynamic limit. This critical behavior, however, is absent in 
related equilibrium spin systems such as the Ising model on BA scale-free networks, 
which in the  thermodynamic limit only displays a ferromagnetic phase. 
The dependence of these results on the degree exponent is also discussed for 
the case of uncorrelated scale-free networks.  
\end{abstract}


\maketitle

\section{Introduction}
Over the last few years, the study of statistical and complex systems has proved 
extremely valuable in providing insight into 
many emerging interdisciplinary fields of science \cite{wei91,wei00,oli99}.  
In particular, many efforts have focused recently on the mathematical modeling 
of a rich variety of social phenomena, 
such as social influence and self-organization, cooperation, 
opinion formation and spreading, evolution of social structures, etc 
(see e.g. \cite{szn00,cas00,ale02,kle03,gon04,suc05,kup06,gon06,can06}). 
In this context, much attention has been devoted to the investigation of social  
models defined on complex networks, since their topology reflects 
some key aspects of social structures, such as the small-world effect and the 
high connectivity of local neighborhoods \cite{wat99,alb02,dor03,new06}. 

The interest in complex networks hugely intensified in recent years due 
to the striking similarity observed in the structure of many 
real networks as different as the Internet and the World Wide Web, ecological and food webs, 
power grids and electronic circuits, genome and metabolic reactions, 
collaboration among scientists and among Hollywood actors, and many others.
 
Empirical observations revealed that the degree distributions of these networks, $P(k)$, 
which measure the probability that any randomly chosen node has $k$ links, 
are fat-tailed and typically close to a power-law, i.e. $P(k)\sim k^{-\gamma}$. Unlike the 
Poisson degree distribution of classical random graphs, power-law distributions lack any 
characteristic scales, hence leading to their designation as {\it scale-free} networks \cite{alb02}. 

Following these intriguing observations, 
Barab\'asi and Albert uncovered the fundamental principles that could be responsible 
for the ubiquitous development of scale-free structures in natural 
and artificial systems \cite{bar99a}. The Barab\'asi-Albert (BA) 
model is based on the assumptions that (i) real networks evolve gradually by the addition of 
new nodes, and that (ii) the mechanism of linear preferential attachment governs the probability 
distribution of connections between the newly added nodes and those already present in the system.  
Indeed, this simple dynamical model for the formation of networks was shown to generate  
structures with scale-free degree distributions with exponent $\gamma_{BA}=3$, 
while detailed explanations for the processes of growth and preferential attachment were provided later 
for several specific cases \cite{alb02,bar99a,bar99b}. 

Besides the extensive and detailed characterization of structural and topological properties of many different 
kinds of complex networks, further investigations have focused on the dynamical and critical behavior 
of models defined on them. As part of these efforts, different spin models 
have been studied on scale-free networks \cite{ale02,igl02,bia02,leo02,dor02,her04,hin06}, 
as well as on other classes of complex networks \cite{can06,bar00,git00,her02,sve02,san02,dor04}. 
On the one hand, 
these investigations have been motivated by the inherent theoretical interest of exploring further properties 
of many standard models of statistical mechanics. On the other hand, spin models defined on complex 
network structures can contribute to the understanding of a rich diversity of processes 
in contexts as different as materials science, sociology, and 
biology. For instance, as regards the application of Ising-type spin models to the study of social phenomena, 
the spin states may denote different opinions or preferences, where the  
coupling constant describes the convincing power between interacting individuals, 
which is in competition with the ``free will'' given by the thermal noise \cite{ale02}. Moreover, 
a magnetic field can be used to add a bias that could be 
interpreted as ``prejudice'' or ``stubbornness'' \cite{sve02}.

Cooperative phenomena arising in models defined on complex networks are profoundly affected by 
the interplay between the model dynamics and the underlying network topology.
For instance, simulations of the Ising model defined on BA scale-free networks reveal that 
the effective critical temperature scales logarithmically with the system size, thus 
showing that only the ordered ferromagnetic phase is present in the thermodynamic limit \cite{ale02}. 
Similar results are obtained for the Ising model defined on uncorrelated scale-free networks with 
exponent $\gamma=3$ \cite{her04}. However, for $2<\gamma <3$ the effective critical temperatures are found to 
scale as a power of the system size, while for $\gamma >3$ the finite-size (pseudo) critical temperatures converge 
to finite values in the thermodynamic limit, hence indicating the occurrence of true ferromagnetic-paramagnetic  
phase transitions \cite{her04}.   

In a related context, nonequilibrium order-disorder phase transitions were recently observed in 
the so-called magnetic Eden model (MEM) \cite{van94,can01} growing on small-world networks \cite{can06}.
The MEM is a kinetic growth model in which the deposited particles have an intrinsic spin. 
In regular lattices, the MEM's growth process leads to Eden-like self-affine growing interfaces and 
fractal cluster structures in the bulk \cite{van94}. Furthermore, the MEM displays 
an interesting variety of nonequilibrium phenomena, such as thermal order-disorder continuous phase transitions and  
spontaneous magnetization reversals, as well as morphological, wetting, 
and corner wetting transitions \cite{can01}.  

The aim of this work is to study the 
irreversible opinion spreading within an inhomogeneous society by characterizing the dynamical 
and critical behavior of the MEM growing on scale-free networks.
According to the growth rules of the MEM, which are given in 
the next Section, the opinion or decision of an individual would be affected by those of their 
acquaintances, but opinion changes (analogous to spin flips in an Ising model) would not occur. 
Moreover, the structure of scale-free networks provides an interesting setting in which the social roles 
differ significantly from one member to another one. Indeed, hubs in the scale-free network 
represent highly influential individuals, whose opinions can potentially affect the decisions of many other 
individuals in the society.  
In order to represent the underlying scale-free topology, we will mainly focus on 
the Barab\'asi-Albert model. However, the dependence of the results on the degree exponent 
will also be discussed by considering the MEM growing on uncorrelated scale-free networks.

Although this work is mainly motivated by social phenomena, a magnetic language will be 
adopted throughout. As commented above, physical concepts such as temperature and magnetization, 
spin growth and clustering, ferromagnetic-paramagnetic phase transitions, etc, 
can be meaningfully re-interpreted in sociological/sociophysical contexts. 

This paper is organized as follows: 
in Sec. II, details on the model definition and the simulation method are given; 
Sec. III is devoted to the presentation and discussion of the results, and Sec. IV
contains the conclusions. 

\section{The model and the simulation method}

According to the model put forward by Barab\'asi and Albert \cite{bar99a}, 
the scale-free nature of many real networks arises from considering two basic dynamical mechanisms: growth and  
preferential attachment. Starting with a small number of nodes, $m$, a new node with $m$ edges   
is added at every time step and is connected to $m$ different nodes already present 
in the system. 
The probability $\Pi$ for connecting the new node to an arbitrary node of degree $k$ is given by  
$\Pi=k/K$, where $K=\sum k$ is the sum of degrees taken over all available nodes in the system. 
Numerical simulations and different analytic studies, such as the continuum theory, the master equation, 
and the rate equation approaches, show that the BA model asymptotically  
leads to scale-free networks with exponent $\gamma_{BA}=3$ \cite{alb02,bar99a,bar99b}.      

After generating one network realization, a randomly chosen up or down spin is deposited on 
a random site. Starting from this seed, the MEM growth takes place by adding, one by one, further spins to the 
immediate neighborhood (the perimeter) of the growing cluster, where the interaction energies between 
neighboring spins shape the growth probability distributions. 
By analogy to the Ising model, the energy $E$ of a configuration of spins is given by
\begin{equation}
E = - \frac{J}{2} \sum_
{\langle ij\rangle} S_iS_j ,
\label{energy}   
\end{equation}
where $S_i= \pm 1$ indicates the orientation of the spin for each occupied site (labeled by the 
subindex $i$), $J>0$ is the ferromagnetic coupling constant between nearest-neighbor (NN) spins, and 
$\langle ij\rangle$ indicates that the summation is taken over all pairs of occupied NN sites (i.e., 
those which are connected within the particular network realization).
As with other spin systems defined on complex networks, the magnetic interaction between any pair 
of spins is only present when a network edge connects their nodes. 

Setting the Boltzmann constant equal to unity ($k_{B} \equiv 1$), the probability 
for a new spin to be added to the perimeter of the existing cluster is
defined as proportional to the Boltzmann factor exp$(-\Delta E /T)$, 
where $T$ is the absolute temperature and $\Delta E$ is the total energy change involved. 
In this work, energy and temperature are measured in units of $J$ throughout. 
At each step, all perimeter sites have to be considered and the probabilities of adding a new (either up 
or down) spin to each site must be evaluated. 
Using the Monte Carlo simulation method, all growth probabilities are first computed and normalized, and then    
the growing site and the orientation of the new spin are both determined by means of a pseudorandom number.
Although the configuration energy of a MEM cluster, given by Eq.(\ref{energy}), resembles the Ising Hamiltonian, it 
should be noticed that the MEM is a nonequilibrium model in which new spins are 
continuously added, while older spins remain frozen and are not allowed to flip. 
The growth of MEM clusters proceeds by iterating this deposition process until the network becomes completely filled. 

Notice that, since all normalized growth probabilities have to be recalculated at each deposition step, 
the resulting update algorithm is quite slow. 
Ensemble averages were calculated by considering typically $10^2-10^3$ different (randomly generated) networks and 
$50-100$ different (randomly chosen) seeds for each network configuration. 
The ensemble-averaging procedures lead to very small statistical errors (typically around or below 
a few percent), which can be safely neglected in the context of qualitative discussions. Unless 
otherwise indicated by means of error bars, statistical errors of data displayed by symbols are 
comparable to or smaller than the symbol sizes.    

The main part of this paper focuses on BA networks, in which different values of the BA parameter 
in the range $1\leq m\leq 7$ are considered. Moreover, results for uncorrelated scale-free networks are 
also presented for 
degree exponents in the range $2\leq\gamma\leq 4$. 
Involving a considerable computational effort, this work presents extensive Monte Carlo simulations 
for BA and uncorrelated scale-free networks with sizes up to $N=2 \times 10^4$. 

\section{Results and discussion}

\begin{figure}[pt]
\begin{center}
\epsfxsize=3.2truein\epsfysize=2.5truein\epsffile{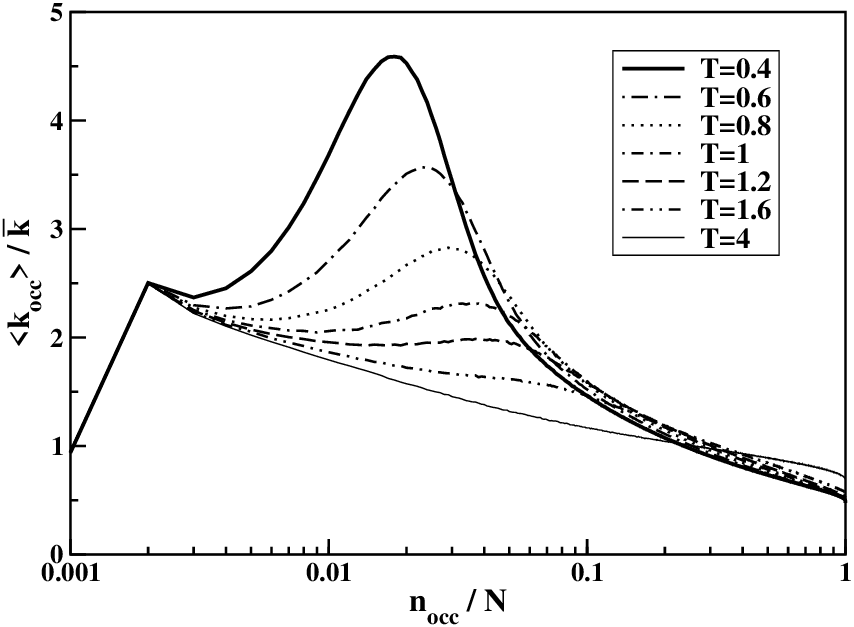}
\caption{Average degree of occupied nodes for BA networks 
of size $N=1000$, $m=3$, and different temperatures, as a function of the 
fraction of deposited particles, $n_{occ}/N$. Notice that, assuming a constant 
deposition rate, the fraction of occupied nodes is a measure of time.
At low temperatures, the early stages of the growth process are dominated by the hubs, 
since the MEM's dynamics favors the occupation of 
highly connected nodes within ordered neighborhoods. At higher temperatures, however, 
the phenomenon of preferential spin deposition is less significant.}
\label{fig1}
\end{center}
\end{figure}

Let us first focus on the dynamical behavior of MEM clusters growing on finite BA scale-free networks. 
In order to investigate the process of spin deposition, we compute the degree of the newly occupied node, $k_{occ}$, 
every time a new particle is added to the system, as well as the number of already occupied NN sites in the neighborhood 
of the growing node, $n_{NN}$, and the energy change involved in the addition of the new spin, $\Delta E$. 

Figure 1 shows the average degree of occupied nodes relative to the network's mean degree, $\langle k_{occ}\rangle / \bar{k}$, 
for BA networks of size $N=10^3$, $m=3$, and different temperatures, as a function of the 
fraction of occupied nodes, $n_{occ}/N$. According to the MEM's growth rules, at low temperatures the system tends to  
develop highly ordered spin domains. Hence, highly connected nodes have larger probabilities to be occupied at early times
during the growth process. The leading role of hubs is clearly observed in the low-temperature 
plots of Figure 1, where $\langle k_{occ}\rangle / \bar{k}\gg 1$ at early stages of the growth process, i.e.  
when the number of occupied nodes is of the order of a few percent of the total system size. At higher temperatures, however, 
the increased thermal noise tends to wash out the phenomenon of preferential spin deposition. The growth at later 
times mainly proceeds by the occupation of less-than-average connected nodes, which leads to roughly temperature-independent 
values of $\langle k_{occ}\rangle$ . 

\begin{figure}[pt]
\begin{center}
\epsfxsize=3.2truein\epsfysize=3.2truein\epsffile{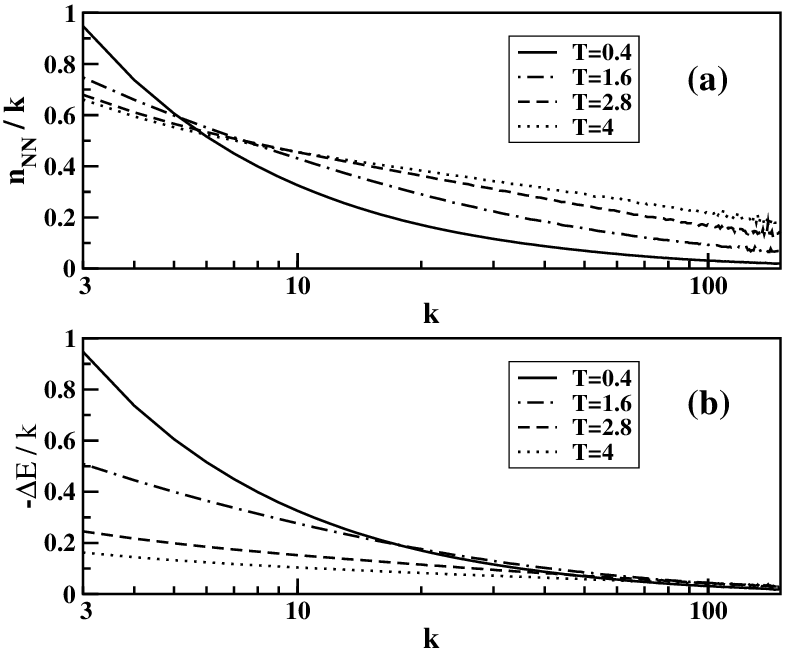}
\caption{Characterization of the deposition process for BA 
networks of size $N=10^3$, $m=3$, and different temperatures, as 
a function of the node degrees: (a) fraction of previously occupied NN sites relative 
to the growing node degree, $n_{NN}/k$; (b) energy change (with opposite sign) involved in the deposition process 
relative to the growing node degree, $-\Delta E/k$.}
\label{fig2}
\end{center}
\end{figure}

Figure 2(a) shows the behavior of the average number of already occupied NN nodes in the local neighborhood of a  
growing node, as a function of its degree, for BA networks of size $N=10^3$, $m=3$, and different temperatures. 
The fraction of occupied NN nodes, $n_{NN}/k$, decreases monotonically with the node degree $k$, 
which is due to the fact that, as discussed above, highly connected nodes are more likely to be occupied at earlier stages of the 
growth process. Moreover, the distributions become flatter for increasingly larger temperatures, as expected.    

In the same vein, Figure 2(b) shows $-\Delta E/k$, i.e. the energy change (with opposite sign) 
involved in the addition of a new spin relative to the degree of the growing site, as  
a function of $k$. In agreement with the previous observations, the plots of $-\Delta E/k$ vs $k$ decrease 
monotonically and become flatter for higher temperatures. 
Notice also that the distributions for the lowest temperature, $T=0.4$, are nearly identical in Figures 2(a) and 2(b), since at low 
temperatures most spins grow preferentially parallel-aligned. Instead, the comparison between plots corresponding to higher 
temperatures shows that $-\Delta E< n_{NN}$, which clearly reflects the onset of thermal disorder.  

Furthermore, it is interesting to note that the energy change involved in the occupation of 
highly connected nodes is very low and roughly temperature-independent. At low temperatures, the hubs become rapidly filled 
and have only a few neighboring spins. Although the number of already deposited neighbors increases for higher temperatures, 
the net local magnetization, as measured by $|\Delta E|$, remains small due to the increased thermal noise.   

\begin{figure}[pt]
\begin{center}
\epsfxsize=3.2truein\epsfysize=2.5truein\epsffile{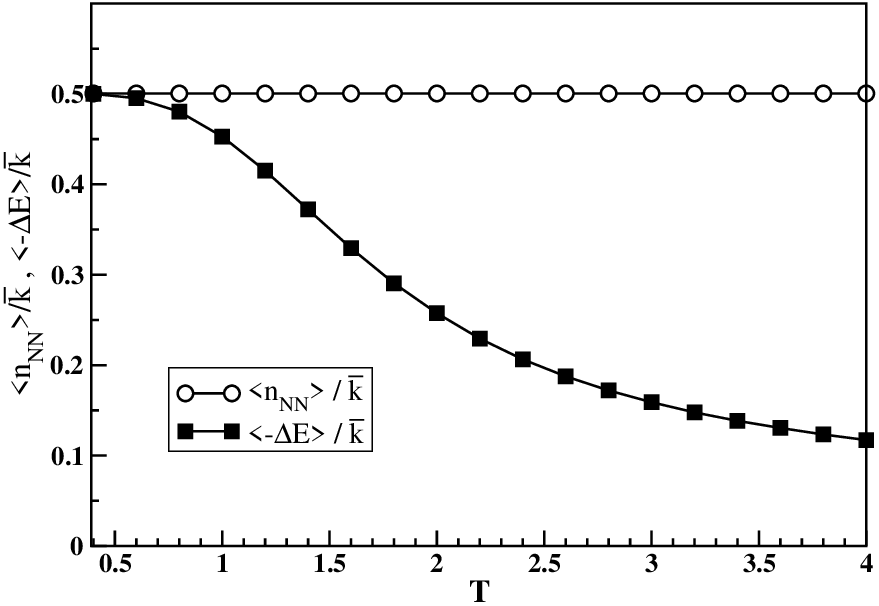}
\caption{Mean number of occupied NN sites and mean energy change involved 
in the addition of a new particle to the system, obtained for BA 
networks of size $N=10^3$ and $m=3$. The temperature-independent 
relation $\langle n_{NN}\rangle=\bar{k}/2$ can be explained by means of a simple bond 
representation that models the growth process (see text for more details).}
\label{fig3}
\end{center}
\end{figure}

Figure 3 shows the mean values of $n_{NN}$ and $\Delta E$ averaged over the full growth process, i.e. until the network 
becomes completely filled, as a function of temperature, for BA networks of size $N=10^3$ and $m=3$. 
At low temperatures, most spins grow parallel-aligned and $\langle -\Delta E\rangle\simeq\langle n_{NN}\rangle$. 
However, at higher temperatures $\langle -\Delta E\rangle$ decreases 
towards the $\langle \Delta E\rangle = 0$ infinite-temperature asymptotic limit, 
while, instead, $\langle n_{NN}\rangle$ remains constant at the value $\langle n_{NN}\rangle=\bar{k}/2$. 

\begin{figure}[pt]
\begin{center}
\epsfxsize=3.2truein\epsfysize=2.5truein\epsffile{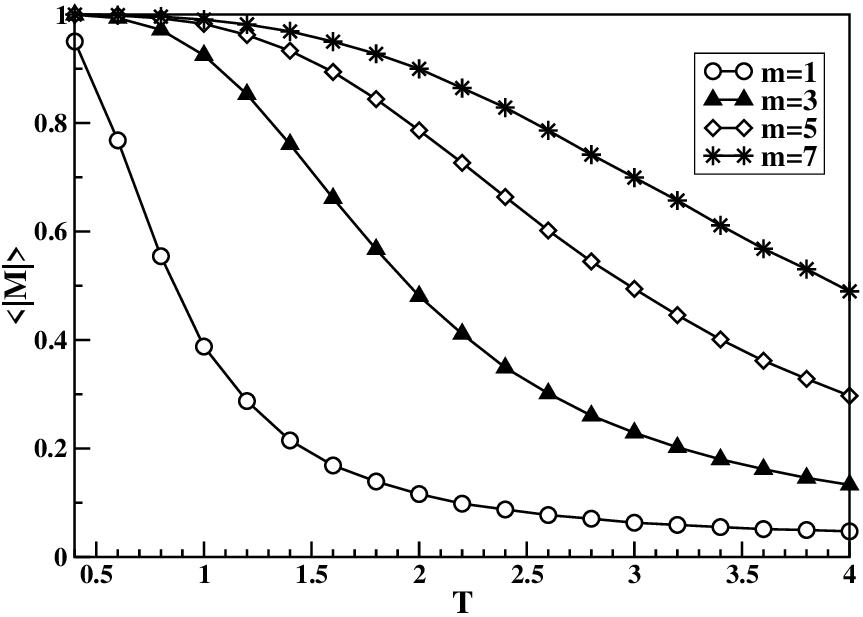}
\caption{Thermal dependence of the order parameter 
for BA networks of size $N=10^3$ and different values of $m$, as 
indicated.}
\label{fig4}
\end{center}
\end{figure}

The latter result can be explained by means of a simple bond representation for the MEM's growth process, as follows. 
To each pair of occupied neighboring nodes, one can assign a directed bond that points from the earlier occupied 
site to the later occupied one. Once the network is completely filled, 
an arbitrary node of degree $k$ will have a net bond flux $\phi$ (i.e. the difference between outwards- and inwards-directed bonds) 
given by $\phi=k-2n_{NN}$. Taking network averages, $\langle\phi\rangle=0$ and thus $\langle n_{NN}\rangle=\bar{k}/2$, 
irrespective of the temperature.  

So far, we have addressed some dynamical aspects of the MEM growing irreversibly on finite-size BA networks. 
Now, let us investigate the critical behavior of this nonequilibrium system. With this aim, we will 
adopt an appropriate order parameter and focus on its thermal dependence and the behavior of associated 
distributions of probability. These results will be later extrapolated to the thermodynamic limit by means of 
finite-size scaling relations. 

The degree of order in a magnetic system can be naturally characterized by the total magnetization per site-i.e.,
\begin{equation}
M={\frac{1}{N}}\sum S_i \ .
\end{equation}
However, according to standard procedures to avoid spurious effects arising from the finite size of the simulated 
samples (see e.g. \cite{oli91,lan00,bin02}), it is more
appropriate to consider instead the absolute value of the total magnetization per site, $|M|$, 
as the order parameter. 

Figure 4 displays the thermal dependence of $\langle |M|\rangle$ for the MEM growing on BA networks of size $N=10^3$ 
and different values of the BA parameter $m$. 
This figure shows, on the one hand, that the effect of increasing $m$ at a fixed temperature is that of 
increasing the net magnetization (and, hence, the order) of the system. Indeed, 
since the mean degree is given by $\bar{k}=m(2-(m+1)/N)\approx 2m$, 
the density of connections grows linearly with $m$ and tends to order the system. 
On the other hand, considering the effect of increasing the temperature for a fixed value of $m$, 
we observe that, at low temperatures, the magnetization is close to unity and the system grows ordered, while, at higher  
temperatures, the disorder sets gradually on and the magnetization becomes significantly reduced. Indeed, this 
behavior reflects the occurrence of thermally-driven, effective order-disorder phase transitions in systems of finite size. 

Further evidence for these pseudophase transitions is obtained from the thermal behavior of the magnetic 
susceptibility, which can be calculated from the fluctuations of the order parameter, as well as from the heat capacity, 
that can be analogously related to energy fluctuations. The thermal distributions of these observables, 
which are not shown here for the sake of space, exhibit peaks that are correlated to the effective critical transitions 
observed in the magnetization plots. 

\begin{figure}[pt]
\begin{center}
\epsfxsize=3.2truein\epsfysize=2.5truein\epsffile{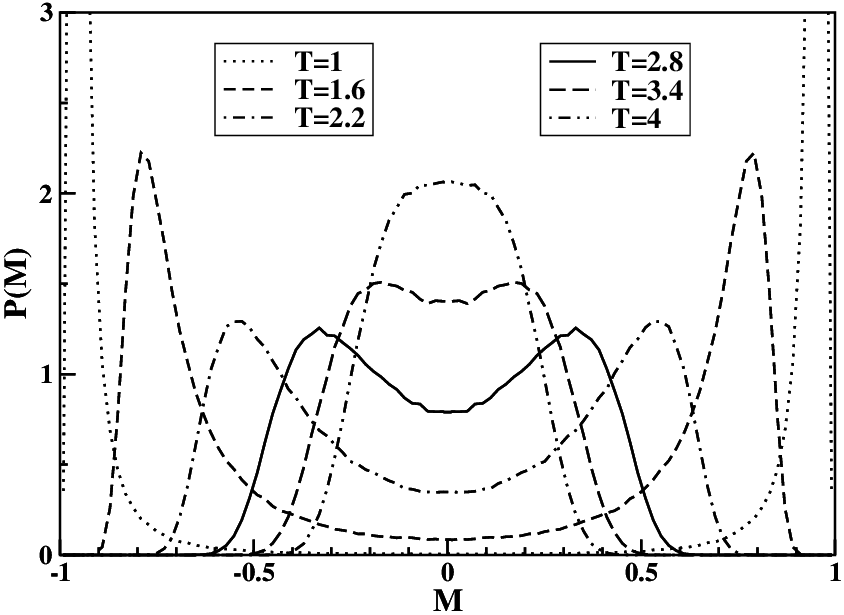}
\caption{Normalized probability distribution of the magnetization, $P(M)$, for BA 
networks of size $N=10^3$, $m=3$, and different temperatures. The sharp 
peaks arising at $M\approx\pm 0.95$ for $T=1$ have been truncated.}
\label{fig5}
\end{center}
\end{figure}

Figure 5 shows the normalized probability distribution of the magnetization, $P(M)$, obtained 
for BA networks of size $N=10^3$, $m=3$, and different temperatures.  
As expected for thermally disordered systems, at high temperatures the distribution is peaked at $M=0$. However, as the 
temperature is lowered, one observes the onset of two maxima located at $M = \pm M_{sp}$ $(0 < M_{sp} < 1)$,
which become sharper and approach $M= \pm 1$ as $T$ is gradually decreased.
The smooth shift of the distribution maxima across $T\simeq T_c$, from the high-temperature $M=0$ 
to the low-temperature nonzero spontaneous 
magnetization $M=\pm M_{sp}$, is the hallmark of true thermally-driven continuous phase transitions \cite{bin02}. 

\begin{figure}[pt]
\begin{center}
\epsfxsize=3.2truein\epsfysize=2.5truein\epsffile{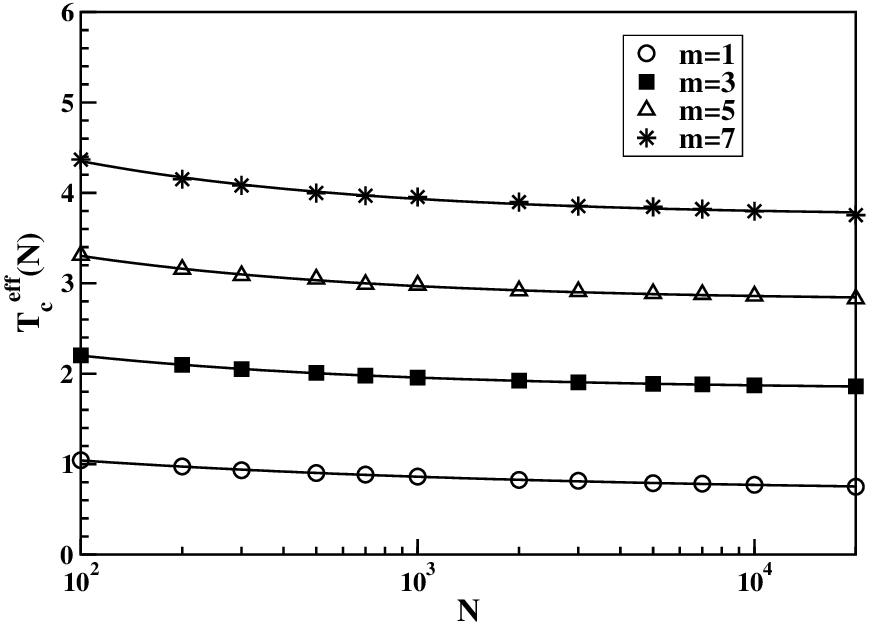}
\caption{Effective transition temperatures for BA networks of different 
size and different values of the parameter $m$ (symbols). Fits to the 
data using the finite-size scaling relation, Eq. (\ref{scaling}), are also 
shown (solid lines).} 
\label{fig6}
\end{center}
\end{figure}

These results provide 
further evidence on the ordering effects induced on dynamical systems by underlying complex network structures, since 
the MEM growing in 1D and 2D regular lattices exhibits only pseudophase transitions with effective critical temperatures 
$T_c^{eff}(N)$ that vanish in the $N\to\infty$ limit \cite{van94}. Moreover,   
the interplay between the topology of BA scale-free networks and the MEM nonequilibrium dynamics leads to 
critical phenomena which are absent in related equilibrium spin systems. For instance, as commented above, 
the Ising model defined on BA scale-free structures shows effective critical temperatures that grow logarithmically with 
the system size, hence displaying only a ferromagnetic phase in the thermodynamic limit \cite{ale02,bia02}. 

The existence of true critical phase transitions, characterized by $m-$dependent finite critical temperatures that separate the 
low-temperature ferromagnetic phase from the high-temperature paramagnetic phase, can be further confirmed by means of a 
standard finite-size scaling procedure. 

According to the finite-size scaling theory, developed for the treatment of finite-size effects 
at criticality and under equilibrium conditions \cite{bar83, pri90},
the difference between the true critical temperature, $T_c$, and an effective pseudocritical 
one, $T_c^{eff}(N)$, is given by 
\begin{equation}
|T_c-T_c^{eff}(N)|\propto N^{-1/\nu}, 
\label{scaling}
\end{equation}
where $\nu$ is the exponent that characterizes the divergence of the correlation length 
at criticality. The effective pseudocritical temperature, $T_c^{eff}(N)$, is here defined as the value corresponding 
to $\langle |M|\rangle=0.5$ for a finite system of $N$ nodes. 

\begin{figure}[pt]
\begin{center}
\epsfxsize=3.2truein\epsfysize=2.5truein\epsffile{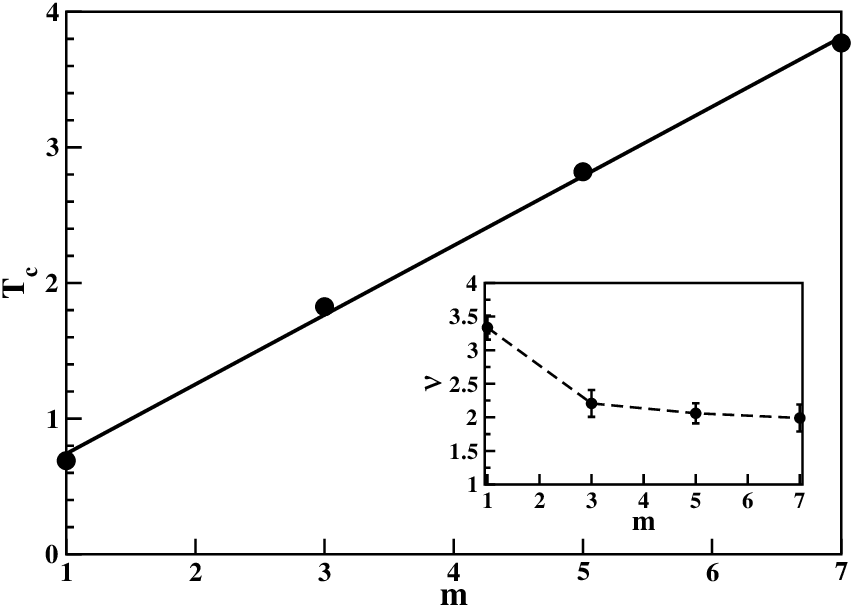}
\caption{Critical temperature as a function of $m$ (symbols), 
which can be approximated by a linear fit (solid line). The inset shows the dependence 
with $m$ of the exponent $\nu$, where the dashed line is a guide to the eye.}  
\label{fig7}
\end{center}
\end{figure}

The symbols in Figure 6 show the effective pseudocritical temperatures for the MEM growing 
on BA networks of different size and different values of the parameter $m$. The solid lines are least-squares 
fits to the data obtained by means of the finite-size scaling relation, Eq. (\ref{scaling}). 
The effective pseudocritical temperatures decrease monotonically with the network size and lead to 
finite extrapolations in the thermodynamic limit, thus confirming the anticipated critical behavior. 
Notice also that, for a fixed system size, the pseudocritical temperatures increase monotonically with $m$, as expected. 

Figure 7 shows the critical temperatures obtained for different values of $m$. The monotonic increase of $T_c(m)$ can be well 
approximated by the linear relation $T_c(m)= 0.522(3)\times m +0.21(1)$, which is displayed in the Figure by a solid line.  
Moreover, the dependence with $m$ of the exponent $\nu$ is shown in the inset, revealing that $\nu$  
decreases monotonically and tends to $\nu\approx 2$ for $m\gtrsim 3$. 

\begin{figure}[pt]
\begin{center}
\epsfxsize=3.2truein\epsfysize=2.5truein\epsffile{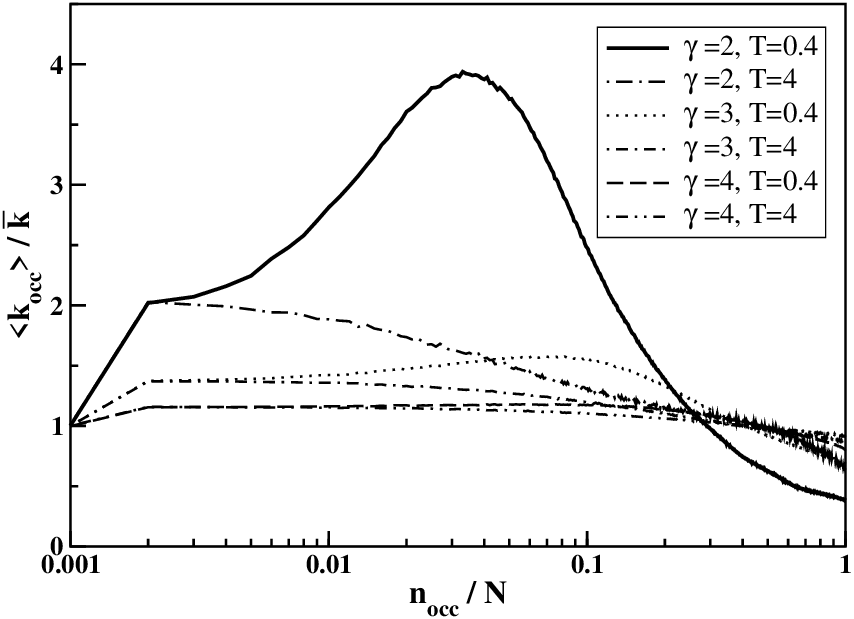}
\caption{Average degree of occupied nodes for uncorrelated scale-free networks 
of size $N\approx 10^3$, minimum degree $k_0=3$, and different values of temperature and 
degree exponent, as a function of the fraction of deposited particles, $n_{occ}/N$. 
The phenomenon of preferential spin deposition, which reveals that hubs dominate 
the early stages of the growth process, is only significant at low temperatures (as 
observed in the case of BA networks, see Figure 1) and for low degree exponents $\gamma<3$.} 
\label{fig8}
\end{center}
\end{figure}

Let us now consider the irreversible growth of MEM clusters on uncorrelated scale-free networks. 
Since this type of networks allows a free choice of the degree exponent $\gamma$, we will investigate 
how the dynamical and critical behavior of the MEM depends on the steepness of the degree distribution.
Moreover, we will compare the critical phenomena observed here to previous investigations in 
a related context, namely the Ising model on uncorrelated scale-free networks \cite{leo02,dor02,her04}.

In order to generate uncorrelated scale-free networks, we assign to each node a given number $k$ 
of ``stubs" (link ends) and match them pairwise at random, with the constraint of avoiding self- and 
multiple connections. 
Besides the exponent $\gamma$, the degree distribution of a finite scale-free network is defined by  
the minimum degree $k_0$ and the upper cutoff $k_{cut}$. Hence, the number of 
sites with degree $k$ is given by $N_k=(k/k_{cut})^{-\gamma}$ for $k_0\leq k\leq k_{cut}$, and $N_k=0$ 
otherwise, while the total number of nodes is $N=\sum_kN_k$ \cite{her04}.  
This procedure, which is based on the so-called configuration model (see e.g. \cite{mol95,mol98,new03}) 
but includes an additional restriction on the maximum possible degree of the vertices, 
was shown to generate scale-free networks with no two- and three-vertex correlations \cite{bog04,cat05}. 

As discussed above, the process of spin deposition during the growth of MEM clusters can be studied 
by computing the degree of the newly occupied node, $k_{occ}$, every time a new particle is added to the 
system (recall Figure 1). Figure 8 shows plots of $\langle k_{occ}\rangle / \bar{k}$ vs $n_{occ}/N$  
for uncorrelated scale-free networks 
of size $N\approx 10^3$, minimum degree $k_0=3$, and different values of $\gamma$, as indicated.  
For each value of the degree exponent considered, we calculated the average degree of occupied nodes for 
two different temperatures that characterize the typical low- and high-temperature behavior, namely 
$T=0.4$ and $T=4$.  
Notice that, due to the procedure followed here to generate uncorrelated scale-free networks, the 
size of the networks is controlled by the degree cutoff $k_{cut}$. Indeed, the actual 
size of the networks represented in Figure 8 fluctuates by $\sim 5\%$ around $N=10^3$. 

The plots in Figure 8 confirm that the phenomenon of preferential spin deposition, 
due to the dominant role played by hubs during the early stages of the growth process, is only significant 
at low temperatures. Moreover, this phenomenon is observed to be relevant only for low degree exponents $\gamma<3$. 
Since the preferential spin deposition is a feature associated to the formation of large ordered clusters during 
the growth process, these results agree well, at a qualitative level, with the analogous behavior reported for the 
Ising model on uncorrelated scale-free networks (see e.g. \cite{her04}), in which the disorder was observed to grow 
monotonically with the exponent $\gamma$.  

Figure 9 displays the thermal dependence of the order parameter, $\langle |M| \rangle$, 
for uncorrelated scale-free networks of size $N\approx 10^3$, minimum degree $k_0=3$, 
and different values of the degree exponent. As discussed above, the degree of disorder in the system 
is observed to grow monotonically with $\gamma$, irrespective of the temperature. Moreover, for any 
fixed value of $\gamma$, the temperature is shown to drive the system through an effective order-disorder 
pseudophase transition. 

\begin{figure}[pt]
\begin{center}
\epsfxsize=3.2truein\epsfysize=2.5truein\epsffile{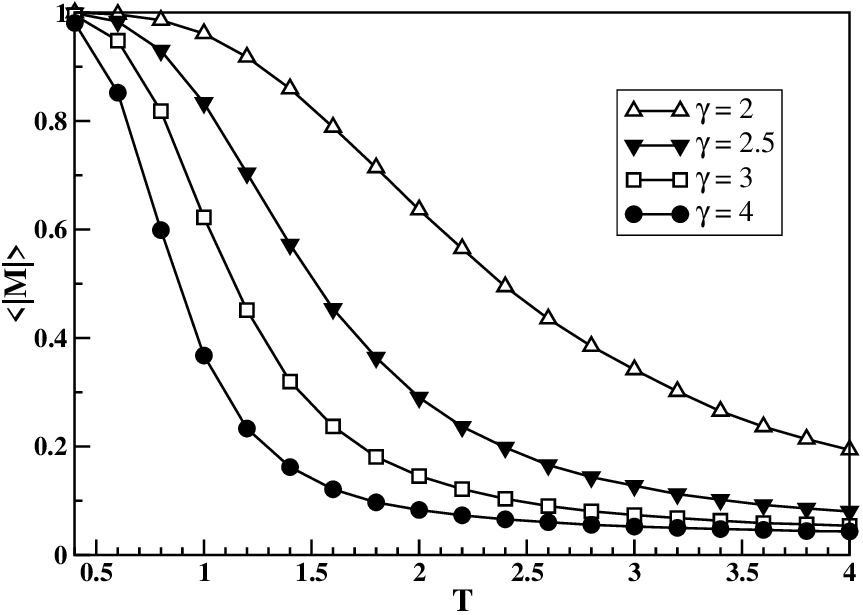}
\caption{Thermal dependence of the order parameter 
for uncorrelated scale-free networks of size $N\approx 10^3$, minimum degree $k_0=3$, 
and different values of degree exponent, as indicated.}
\label{fig9}
\end{center}
\end{figure}

\begin{figure}[pt]
\begin{center}
\epsfxsize=3.2truein\epsfysize=2.5truein\epsffile{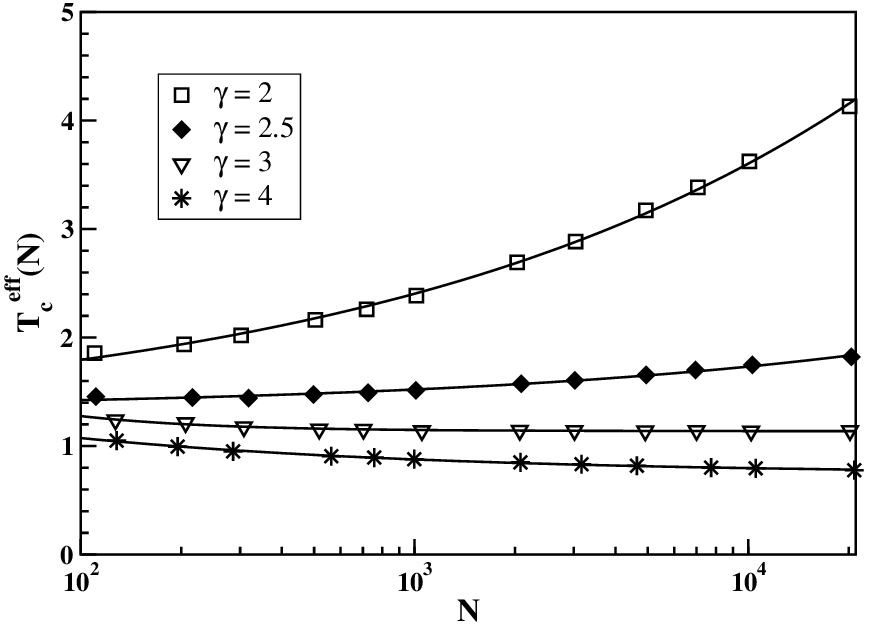}
\caption{Effective transition temperatures for uncorrelated scale-free networks of 
minimum degree $k_0=3$, different 
size, and different values of degree exponent (symbols). Fits to the 
data using the scaling relations given by Eqs. (\ref{scaling}) and (\ref{scaling2}) 
are also shown (solid lines).}
\label{fig10}
\end{center}
\end{figure}

In order to further investigate the scaling properties of the observed pseudocritical 
behavior, we can determine the effective pseudocritical temperatures for systems of different size and  
extrapolate the results to the thermodynamic limit. Analogously to the procedure followed above (see Figures 6 and 
7), symbols in Figure 10 represent the effective transition temperatures for the MEM growing on uncorrelated scale-free networks of 
minimum degree $k_0=3$, different size, and different values of the degree exponent. 
The critical behavior turns out to be crucially dependent on the steepness of the degree distribution. 
Indeed, while the plots 
with $\gamma\geq 3$ look similar to those obtained previously for the MEM growing on BA networks, the plots corresponding 
to smaller values 
of the degree exponent are observed to diverge, hence implying the absence of paramagnetic-ferromagnetic phase transitions in 
the thermodynamic limit. 

Using the finite-size scaling relation given by Eq. (\ref{scaling}), the plots for $\gamma\geq 3$ can be well fitted 
(solid lines in Figure 10). For $\gamma=3$, we obtain $T_c=1.14(2)$ and $\nu=0.9(1)$, while, for $\gamma=4$, the 
scaling relation yields $T_c=0.74(1)$ and $\nu=2.6(2)$. 

In the same vein, the divergent plots for $\gamma<3$ can be well fitted by a scaling relation analogous to Eq. (\ref{scaling}), 
namely 
\begin{equation}
T_c^{eff}(N) = A + B\times N^z\ , 
\label{scaling2}
\end{equation}
as shown by the corresponding solid lines in Figure 10. The $\gamma=2$ plot diverges with exponent 
$z=0.29(3)$, while, for $\gamma=2.5$, one obtains $z=0.33(3)$.   

Comparing these observations with the results reported for the Ising model on uncorrelated scale-free 
networks \cite{leo02,dor02,her04}, we find a qualitative agreement for both $\gamma<3$ (i.e. the absence of 
a paramagnetic phase in the thermodynamic limit) and $\gamma>3$ (i.e. the existence of a finite critical temperature 
delimiting the paramagnetic-ferromagnetic phase transition). 
Instead, the convergence of the critical temperature for the $\gamma=3$ case is clearly in contrast with the logarithmic divergence 
observed in the Ising model. Indeed, this was also pointed out above when comparing the behavior of both models defined on BA 
structures. 
We can thus stress the fact that, for both the MEM and the Ising model, the critical 
features remain unchanged when passing from the BA network topology to the uncorrelated scale-free network structure 
with degree exponent $\gamma=3$.  

\section{Conclusions}

Dynamical and critical properties of the magnetic Eden model growing on Barab\'asi-Albert scale-free networks 
were studied by means of extensive Monte Carlo simulations. According to former studies on spin models defined on 
complex networks, magnetic concepts such as e.g. temperature, spin states, and ferromagnetic couplings, can be meaningfully 
reinterpreted in social contexts. Hence, the MEM growing on scale-free networks is here proposed as a model for the 
irreversible opinion spreading within an inhomogeneous society, in which different individuals play different 
social roles. 

The deposition/spreading dynamics was studied as a function of the degree of the occupied 
sites and showed evidence for the leading role played by hubs in the growth process. Indeed, 
it was observed a phenomenon of preferential spin deposition, in which highly connected nodes 
tend to dominate the growth dynamics at early times. 
This phenomenon is more significant at low temperatures, since the 
MEM's growth rules favor the occupation of highly connected nodes within ordered local neighborhoods. 

Adopting the mean absolute magnetization per spin as the order parameter, we observed the occurrence 
of effective pseudocritical order-disorder phase transitions. The thermal dependence of the order 
parameter distribution functions showed the characteristic signatures of true thermally-driven 
continuous phase transitions, which were indeed confirmed by means of standard finite-size 
scaling procedures.     

The critical behavior found in this nonequilibrium spin model contrasts remarkably with the behavior 
observed in related equilibrium spin systems such as the Ising model on BA scale-free networks, 
which in the thermodynamic limit only displays a ferromagnetic phase. 

Finally, the irreversible growth of MEM clusters was studied on uncorrelated scale-free networks. 
Many of the qualitative features in the growth of finite samples were found to be common to 
both BA and uncorrelated scale-free networks. However, the latter were observed to exhibit a very strong dependence on 
the degree exponent $\gamma$, which manifests itself most significantly in the critical properties of the system. 
The critical behavior for $\gamma\geq 3$ was found analogous to 
previous results for the MEM growing on BA networks, while, for smaller values 
of the degree exponent, only the ferromagnetic phase was observed to persist in the thermodynamic limit. 
 
The present findings will thus hopefully stimulate and contribute to the growing interdisciplinary efforts in 
the fields of sociophysics, complex networks, and nonequilibrium statistical physics.  

\section*{Acknowledgments}
This work was supported by the James S. McDonnell Foundation and the National 
Science Foundation under grants INT-0336343, ITR DMR-0426737, and CNS-0540348 within the DDDAS 
program. The author thanks Albert-L\'aszl\'o Barab\'asi for useful discussions.

\end{document}